\documentstyle[12pt,aaspp4]{article}

\received{2001 August 22}

\newcommand{\ep}{$e^-e^+$}

\begin{document}

\title{Formation and dynamics of self-sustained neutron haloes 
in disk accreting sources} 

\author{A.A. Belyanin$^{1,2}$ and E.V. Derishev$^{1,3}$}

\affil{
$^1$Institute of Applied Physics, Russian Academy of Science \\
46 Ulyanov st., 603950 Nizhny Novgorod, Russia\\
$^2$Dept. of Physics, Texas A\&M University, College Station,
TX 77843-4242, USA\\
$^3$MPI f\"{u}r Kernphysik, Saupfercheckweg 1, D-69117 Heidelberg, Germany}

\authoremail{belyanin@atlantic.tamu.edu, derishev@mpi-hd.mpg.de}

\date{}

\begin{abstract}

It has been recognized long ago that the presence of hot plasma 
in the inner accretion disks around black
holes could lead to the neutron  production via 
dissociation  of  helium  nuclei. 
We show that, for a broad range of accretion parameters,
neutrons effectively decouple from protons and pile up in the inner disk
leading to the formation of  self-sustained halo.  This means that new 
neutrons in the halo are supplied mainly by the splitting of helium nuclei 
in their collisions with existing neutrons. Once formed, such a halo can
exist even if the proton temperature is much lower than the energy 
threshold of helium dissociation. 
We show that neutron haloes  can be the natural source of relativistic
electrons and positrons, providing characteristic comptonization spectra
and  hard spectral tails observed in many black hole candidates, and also
giving rise to relativistic outflows. Deuterium gamma-ray line at
2.2 MeV resulting from neutron capture is also expected at a level
detectable by future INTEGRAL mission.   Furthermore, the presence of
a neutron halo strongly affects the dynamics of accretion and leads to the
rich variety of transient dynamical regimes.  


\end{abstract}

\newpage

     \section{Introduction}

    There are many theoretical and observational indications that
the hot plasma of ion temperature $T_i$ exceeding several MeV can
exist  in  the inner accretion disks around black holes and, possibly, neutron stars. Formation
of  a two-temperature  region with $T_i \gg T_e$ in accretion disks
has  been  predicted  as long ago as in 1976 (Shapiro et al. \cite{s76}). When the
energy  transfer  from ions to electrons is slow as compared with
the accretion timescale, an accretion flow becomes nearly adiabatic,
and  the  ion  temperature can reach virial values. This idea has
led  to  the  prediction  of  advection-dominated accretion flows
  which  seems  to find support in observations of several
X-ray  transients (Narayan et al. \cite{N96}, \cite{N97}) and other sources showing low 
radiation efficiency and high temperature of the flow. 

    Strong   observational   support   for   the   existence   of
high-temperature plasma comes from gamma-ray and radio astronomy.
Very  hard  spectra  extending  to  MeV region were detected from
several  Galactic X-ray sources presumably containing black holes.
  Transient line-like features around 0.5 MeV indicate
an   efficient   mechanism   of   positron  production  (Bouchet et al. 
\cite{Sun1}, Churazov et al. \cite{Sun2}).
Non-thermal  radio emission and episodic ejection of relativistic
jets  observed in microquasars (Mirabel and Rodr\'{i}guez \cite{MR}) are definite signatures of
ultrarelativistic electrons and/or electron-positron (\ep) plasma
and    were    interpreted    several    times    as    
radiation-pressure dominated \ep\ outflows (Li and Liang \cite{LL96}, 
Belyanin \cite{B99}).

    The presence of hot ions in accretion disks leads to the rich
variety  of inelastic nuclear interactions. The most prominent of
them   is  neutron  production  which  primarily  occurs  through
dissociation  of  helium  nuclei.  This  process  has potentially
important    observational   implications   arising   e.g.   from
possibility   of   radiative   neutron  capture  with  subsequent
formation  of  gamma-ray  lines  (Aharonyan and Sunyaev \cite{AS}).

The main source of free neutrons in accretion disks is dissociation
of helium nuclei due to inelastic collisions with energetic ions
contained in the disk or free neutrons passing through it. 
Neutron  production in
accretion  disks  was  considered several times (Aharonyan and Sunyaev \cite{AS}, Guessoum and Kazanas \cite{G89}) under the
assumption  that neutrons are quickly thermalized in the disk, advected 
by the bulk plasma motion and transported into a compact object on the 
disk viscous timescale. In this
case  the  neutron  fraction  in the accretion flow is inevitably
small  and  the  radiative-capture  gamma-ray  lines are weak and
smeared  out  by a  hot  rapidly  rotating disk. 

However, if plasma 
in the accretion disk is sufficiently rarefied, then the neutron and 
ion components in the disk effectively decouple, allowing for the 
accumulation of neutrons. This can lead to the formation of neutron 
haloes with much larger neutron number than in previously considered cases. 
It  is  important that, once
formed,  such  a  halo  becomes  {\it self-sustained}. This means
that,  even  if  the  ion temperature in the disk falls below the
threshold  for He dissociation by protons or the proton density
becomes too small for proton-induced dissociation to be efficient, 
the neutron production
is supported by collisions of energetic neutrons from the halo
with helium nuclei. 

In this paper we propose and explore the scenario of neutron halo
formation
and show that it can lead to much more pronounced observational  and
dynamical
effects as compared with previous  predictions.

\section{Neutron pile-up and formation of halo}

Let us consider bulk radial motion of neutrons which is the
result of angular momentum losses caused by elastic collisions with
ions (mostly protons) in the accretion disk. One directly obtains the
following expression:
\begin{equation}
V_n^{(r)} =
- \frac{d}{dt} \left( \frac{R_g}{2} \frac{c^2}{V_n^2} \right)
= R_g \frac{c^2}{V_n^3} \frac{d V_n}{dt} =
- \frac{R_g}{2} \frac{c^2}{V_n^3} \nu_{pn} \left( V_n - V_d \right)
\, ,
\end{equation}
where $V_n$ and $V_n^{(r)}$ are the orbital and radial velocities of
neutrons, $R_g$ the Schwarzschild radius of the compact object, and
$c$ the speed of light. The time derivative of $V_n$ in the above
expression was substituted by $- \nu_{pn} (V_n - V_d)/2$, where
$\nu_{pn}$ is the proton-neutron collision rate and $V_d$ the disk
orbital velocity.

As the orbital velocity of neutrons is larger than that of the
accretion disk, the excessive centrifugal force has to be balanced by
the friction force, which originates from the difference of radial
velocities of the neutrons and the disk. This gives another relation:
\begin{equation}
\frac{V_n^2 - V_d^2}{R} =
 \frac{\nu_{pn}}{2} \left( V_n^{(r)} - V_d^{(r)} \right)\, .
\end{equation}
Here $V_d^{(r)}$ is the radial velocity of the disk and $R$ the
current orbital radius. Assuming $V_n \simeq V_d \simeq V_0$ ($V_0$ is
the Keplerian velocity), one finds the expression for the radial
velocity of neutrons:
\begin{equation}
V_n^{(r)} = - \frac{1}{4} \left( \frac{R}{V_0} \nu_{pn} \right)^2
\left( V_n^{(r)} - V_d^{(r)} \right) .
\end{equation}
The value $(R/V_0) \nu_{pn}$ in the above equation is the number
of collisions a neutron undergoes as it completes one orbital
revolution, divided by $2\pi$.

The collision rate $\nu_{pn}$ may be expressed in terms of the disk
optical depth for proton-neutron collisions, $\tau_d$, which is equal
to the average number of collisions experienced by a neutron passing
through accretion disk with Keplerian velocity directed perpendicular
to the disk plane. Given the heights of ion and neutron "disks",
$h_d$  and $h_n$ respectively, and assuming that the cross-section of
proton-neutron collisions is inversely proportional to their relative
velocity, one has $\nu_{pn} = \left< \sigma V \right>_{pn} n_p
(h_d/h_n) = (V_0/h_n) (\left< \sigma V \right>_{pn} /V_0) n_p h_d =
(V_0/2h_n) \tau_d$. Finally, the expression for the radial velocity of
neutrons takes the following form:
\begin{equation}
V_n^{(r)} = \frac{(R/2h_n)^2 \tau_d^2}{4 + (R/2h_n)^2 \tau_d^2}
V_d^{(r)}\, .
\end{equation}
Neutrons start to pile up in the inner parts of accretion disk when
$V_n^{(r)}$ is considerably smaller than $V_d^{(r)}$, i.e., when
$\tau_d \la 4 h_n/R$. In the limiting case of neutron halo
($2 h_n \sim R$) the last condition gives $\tau_d \la  1$.

\section{Criteria of neutron halo existence}

The neutrons in accretion disk pile up only when the accretion rate
$\dot{M}$ is not too large, namely
\begin{equation}
\dot{M} \la \dot{M}_{max} =
\frac{8 \pi V_d^{(r)} V_0 h_n}
{\left< \sigma V \right>_{pn}} m_p =
\frac{4 \pi \alpha R_g c^2}{\left< \sigma V \right>_{pn}}
\left( \frac{h_d}{R} \right)^{2} \left(\frac{h_n}{R}\right) m_p \, .
\end{equation}
Here we used the usual assumption $V_d^{(r)} = \alpha (h_d/R)^2 V_0$,
and the continuity equation to find the proton number density. 
The ratio $h_d/R$ is treated as a parameter. The limiting accretion
rate $\dot{M}_{max}$ approximately equals to that corresponding to
the Eddington luminosity, provided $\alpha = 0.1$ and $h_d \simeq R$.

With decreasing accretion rate the neutron-to-proton ratio in the
disk grows as $\dot{M}^{-2}$ and soon exceeds unity, so that
collisions with neutrons become the main cause of helium
dissociation. Actually, it happens at a lower neutron-to-proton
ratio, because the neutron component tends to have higher temperature
than the ion one. Even in the absence of additional mechanisms that
act specifically to heat neutrons (e.g., precession of neutron's
orbit in the gravimagnetic field of a rotating compact object), their
temperature exceeds the ion temperature by a factor 7/5.

In view of the above-mentioned, the most interesting is the
self-sustained mode of neutron halo formation, when new neutrons in
the halo are supplied solely by the splitting of helium nuclei in
their collisions with existing neutrons. This mode implies that at
least one of three neutrons breaks helium nucleus before decaying or
being advected into the black hole. These two conditions provide,
respectively, lower and upper bounds on the range of accretion rates
in which the self-sustained neutron halo may exist. 

Let us consider the upper boundary first. To do this, we calculate the
average number of collisions with protons experienced by each neutron
on its way into the black hole:
\begin{equation}
N_c = \frac{\nu_{pn} R}{V_n^{(r)}} =
\frac{2 \pi R_g c^2}{\left< \sigma V \right>_{pn}\dot{M}}
\left( 4 + \left[ \frac{R}{V_0} \nu_{pn}
\right]^2 \right) \frac{h_n}{R} m_p\, ,
\end{equation}
assuming zero helium mass fraction $\eta_{He}$. In the following
considerations we neglect the term $({R}/{V_0}) \nu_{pn}$ as it is
important only if $\dot{M} \simeq \dot{M}_{max}$. The self-sustained
neutron halo requires
\begin{equation}
N_c > \frac{1}{3} \frac{4}{\eta_{He}} \frac{\sigma_{pn}}{\sigma_d}
\quad \Rightarrow \quad
\dot{M} < \frac{3 \eta_{He}}{2 \alpha}
\frac{\sigma_d}{\sigma_{pn}}
\left( \frac{R}{h_d} \right)^2 \dot{M}_{max}\, ,
\end{equation}
where  $\sigma_{pn}$ the proton-neutron scattering
cross-section, having the value $\sigma_{pn} \simeq 700$~mb at the 
energies typical for the edge of neutron halo, and $\sigma_d \simeq 100$~mb 
is the dissociation cross-section in $n-\alpha$ collisions which, 
in effect, is the sum of cross-sections for all possible spallation 
processes. It is clear that the self-sustained
mode maintains up to $\dot{M} \simeq 0.4 \dot{M}_{max}$ or even
higher; at still higher accretion rates inelastic $p-\alpha$
collisions are frequent enough to ensure complete dissociation of
helium.

At very low values of $\dot{M}$ the neutrons may be advected so slowly
that their finite lifetime, $t_n \simeq 900$~s, comes into play.
The decay losses are inevitable (and irreversible), but they become a
limiting factor only if there are no other processes to compete with,
for example, evaporation of neutrons from the disk. This means that the
ion temperature should be considerably lower than the virial value
or, in other words, $h_d \ll R$. Keeping in mind this restriction,
one may derive the absolute lower limit on the accretion rate (the one that
still permits the self-sustained neutron halo) from the condition
\begin{equation}
\frac{1}{4} \eta_{He} \frac{\sigma_d}{\sigma_{pn}} \nu_{pn} t_n
> \frac{1}{3},
\end{equation}
which gives
\begin{equation}
\label{Mmin}
\dot{M} > \dot{M}_{min} =
\frac{16 \pi}{3} \frac{\sigma_{pn}}{\sigma_d}
\frac{R h_n V_d^{(r)}}
{\eta_{He} \left< \sigma V \right>_{pn} t_n} m_p =
\frac{4}{3\sqrt{2}} \frac{\sigma_{pn}}{\eta_{He} \sigma_d}
\frac{R_g}{c t_n}
\left( \frac{R}{R_g} \right)^{3/2} \dot{M}_{max} \, .
\end{equation}

The region occupied by the neutron halo is limited in extent by the
requirement that virial temperature at the edge of the halo
approaches helium dissociation threshold $\simeq 20$~MeV. The latter means 
$R/R_g \la 30$. Substituting this ratio in Eq. (\ref{Mmin}) one finds
$\dot{M}_{min} \sim 10^{-3} \dot{M}_{max}$ for a 
black hole of ten solar masses and $\eta_{He} = 0.2$. Thus, there is a 
large parameter
space in which the neutron halo may exist. It should be noted that the
halo can be formed even if $\dot{M} < \dot{M}_{min}$, but with a 
smaller radius. In principle, this
marginal solution could exist up to the accretion rates as low as
$\sim 0.03 \dot{M}_{min}$, but in this case it faces a serious threat
from rapidly increasing neutron losses in the immediate vicinity of
the last stable orbit. This regime, as well as the dynamics of neutron 
halo in the limit of prevailing neutron-neutron collisions, will be 
considered elsewhere.  

\section{Virialization of neutrons in the halo}

As was mentioned above, precession of neutron's orbit caused by
the gravimagnetic field of a rotating black hole results in additional
heating of neutrons and may lead to complete virialization of
the neutron halo if the precession frequency is higher than or of the
order of one half of the neutron-proton collision rate, i.e.,
\begin{equation}
\frac{ac}{2\pi R_g} \left( \frac{R_g}{R} \right)^3
\ga \frac{\nu_{pn}}{2},
\end{equation}
where $a$ is the black hole angular momentum in units $M R_g c/2$.
Substituting $\nu_{pn}$ in the above equation it is easy to find the
corresponding inequality for the accretion rate:
\begin{equation}
\label{virial}
\dot{M} \la \frac{a}{\sqrt{2} \pi}
\left( \frac{R_g}{R} \right)^{3/2} \dot{M}_{max}\, .
\end{equation}
Even for a rapidly rotating ($a \rightarrow 1$) black hole Eq.
(\ref{virial}) requires mass accretion rate hardly consistent with
the lower limit $\dot{M}_{min}$. Therefore, we conclude that the
mechanism under consideration is unlikely to cause virialization of
the neutron halo close to its outer edge. However, the virialization is
possible in the inner part of the halo at the distance
\begin{equation}
\label{virial2}
R \la
\left( \frac{a}{\sqrt{2} \pi} \frac{\dot{M}_{max}}{\dot{M}} \right)^{2/3}
R_g\, .
\end{equation}

\section{Long-term variability}

As the neutron halo close to the black hole is likely to be
virialized, it produces a neutron wind which is nearly isotropic,
expands with a velocity comparable to the velocity of light, and
 may carry about a half of all neutrons extracted from helium in an
accretion disk.  Note that helium nuclei completely dissociate when
the accretion rate is capable to support a self-sustained neutron
halo or is comparable with $\dot{M}_{max}$, i.e. at any $\dot{M} \ga
\dot{M}_{min}$. The neutron wind exerts a dynamical pressure on
the infalling material via the protons originating from the neutron
decay. At a distance smaller than $t_n V_w$ ($V_w$ is a typical
velocity of neutrons in the wind) this pressure may be expressed in
the following form
\begin{equation}
P_w \simeq \frac{\eta_{He}}{4} \frac{\dot{M}}{4\pi R t_n}.
\end{equation}
In the case of quasi-spherical accretion (e.g., from stellar wind 
emanating from a massive companion star) the dynamical pressure of the 
neutron
wind prevents the surrounding matter from infall provided the dynamical
pressure of an accreted plasma at the sonic point $R_s \simeq
(c/V_{\infty})^2 R_g$ ($V_{\infty}$ is the velocity of the black hole
with respect to the stellar wind is less than $P_w$. 
This gives the condition
\begin{equation}
\label{pressure}
\frac{\dot{M} V_{\infty}}{4\pi R_s^2} \la P_w
\quad \Rightarrow \quad
\frac{V_{\infty}}{c} \la
\left( \frac{\eta_{He}}{4} \frac{R_g}{c t_n} \right)^{1/3}
\end{equation}

As soon as the condition (\ref{pressure}) is satisfied, the accretion
stops, but the source remains active thanks to the matter accumulated
in an accretion disk. The disk becomes exhausted in a time approximately
equal to its viscous timescale at the outer radius, and after that
the source switches off allowing for a new cycle of accretion from
a stellar wind. The quiet period lasts for roughly the same time as the 
active one, until the matter initially deposited at the edge of the
disk reaches the neutron halo boundary. Such a mechanism results in
a long-term variability with a period very sensitive to the disk
outer size, i.e. to the angular momentum per unit mass of the
infalling material.  The maximum period may be hundreds or
thousands years.

\section{Proton-neutron bremsstrahlung}

Some of the proton-neutron collisions result in the emission of
gamma-quanta, either in deuterium formation process or due to
bremsstrahlung, first suggested to be an efficient source of radiation
in proton-neutron plasma by Aharonyan and Sunyaev (\cite{AS}). 
Let us define $\sigma_{\gamma}$ so that the energy
radiated in each collision of a neutron with a proton constitutes on
average $\sigma_{\gamma}/\sigma_{pn}$ fraction of their relative
kinetic energy. The typical value of this ratio is of $\sim 10^{-4}$ in our case. 
Therefore, the energy radiated by a neutron per unit radius is
\begin{equation}
\frac{d\epsilon_n}{dR} =
\frac{\sigma_{\gamma}}{\sigma_{pn}} (3T + \Delta) \frac{dN_c}{dR}.
\end{equation}
Here $\Delta \simeq 2.2$~MeV is the binding energy of deuteron, and
$T$ is an average temperature of protons and neutrons.
The radiative efficiency of a disk  due to proton-neutron bremsstrahlung, 
as follows from the above equation, is equal to
\begin{equation}
\frac{(\eta_{He}/2) d\epsilon_n/dR}{(R_g/4R^2) m_p c^2} =
\eta_{He} \frac{\sigma_{\gamma}}{\sigma_{pn}} (3T + \Delta)
\frac{16\pi h_n}{\left< \sigma V \right>_{pn}\dot{M}}.
\end{equation}
This efficiency approaches unity when
\begin{equation}
\dot{M} \la 8\pi \eta_{He} \frac{\sigma_{\gamma}}{\sigma_{pn}}
\frac{R_g c^2}{\left< \sigma V \right>_{pn}} \frac{h_n}{R} m_p =
\frac{2 \eta_{He}}{\alpha} \frac{\sigma_{\gamma}}{\sigma_{pn}}
\left( \frac{R}{h_d} \right)^2 \dot{M}_{max}\, ,
\end{equation}
assuming $\Delta \ll 3T$ and $3T \simeq (R_g/2R) m_p c^2$. At an
accretion rate exceeding the above limit, luminosity due to
proton-neutron bremsstrahlung remains constant independently on the
accretion rate.

\section{Pion production and hard gamma-ray luminosity of the halo}

As was mentioned above, the inner part of the neutron halo is most
likely to be virialized. Therefore, neutron-proton and neutron-neutron
collisions within $R \sim 3 R_g$ lead to an  efficient production of
pions. If we assume for simplicity that the mean pion-production rate
$\xi_{\pi} = \langle \sigma_{\pi} V\rangle$, averaged over the
velocity distribution of neutrons, is the same for both kinds of
collisions, the total luminosity in pions is
\begin{equation}
\label{pi1}
L_{\pi} \sim \xi_{\pi} (\frac{1}{2} n_n^2 +
\frac{h_d}{h_n} n_n n_p) R_{\pi}^3 E_{\pi},
\end{equation}
where $R_{\pi}$ is the radius within which the pions are produced,
$E_{\pi} \simeq 140$ MeV is the pion energy.  If He dissociation is
complete, the density of neutrons in the inner halo can be found from
the following  balance condition:
\begin{equation}
\label{pi2}
\frac{\eta_{He} \dot{M}}{2 m_p} =
\xi_{\rm loss} \frac{n_n^2}{2} R_{\pi}^3,
\end{equation}
where the rate of neutron losses $\xi_{\rm loss} $ is mainly determined by the scattering  into a black hole.
Using the expression for $n_n$ from Eq.~(\ref{pi2}), the expression
for $n_p$ derived from the mass continuity equation, and
assuming $h_n = R$, we obtain pion luminosities associated with
neutron-neutron and neutron-proton collisions:
\begin{equation}
\label{pi3}
L_{nn} = \frac{\xi_{\pi}}{\xi_{\rm loss}} \frac{\eta_{\rm He} \dot{M} E_{\pi}}{2 m_p},
\end{equation}
\begin{equation}
\label{pi4}
L_{pn} = \frac{E_{\pi}}{2\pi \alpha} \frac{\xi_{\pi}}{\xi_{\rm loss}}
\left( \frac{R}{h_d}\right)^2 \left( \frac{\eta_{\rm He} \xi_{\rm
loss} \dot{M}^3}{R_g c^2  m_p^3}\right)^{1/2}.
\end{equation}
The upper limit on luminosities can be obtained by putting $\dot{M} =
\dot{M}_{\rm max}$.  This gives
\begin{equation}
\label{pi5}
L_{nn}^{\rm max} \simeq 7\times 10^{38} \alpha \eta_{\rm He} M_{10}
 \left(\frac{h_d}{R}\right)^2 \; {\rm erg/s},
\end{equation}
\begin{equation}
\label{pi6}
L_{pn}^{\rm max} \simeq 10^{38} M_{10}\sqrt{\alpha \eta_{\rm He}}
\left(\frac{h_d}{R}\right) \; {\rm erg/s},
\end{equation}
where $M_{10}$ is the black-hole mass in units of 10 solar masses.
Approximately 1/2 of the total pion luminosity goes to gamma-quanta
and \ep\ pairs and can be transformed to radiation via inverse
Compton scattering of soft X-rays and electromagnetic cascade.

In the presence of abundant soft X-rays from disk, relativistic \ep\
pairs are quickly cooled, and the snapshot of electron distribution
would demonstrate that the average energy is between
50-500 keV. The resulting photon spectrum above 100 keV has a typical
comptonization pattern. Pair cascade, if developed, leads to a hard
power-law tail with photon index around 2. Both features are
frequently observed in the hard state of accreting black-hole
candidates. In addition, very hard radiation should be present with
photon energies up to 70 MeV, but it is expected to contain much
less energy than hard X-rays and is difficult to observe.

Neutron halo can be also an efficient source of  deuterium line at
$\simeq 2.2$ MeV. It originates from radiative capture of neutrons
that evaporate from the halo and are intercepted by the cold outer disk.  The total
luminosity in the line is $\beta \eta_{\rm He} \dot{M} \Delta/(2
m_p)$, where $\beta$ is the fraction of neutrons that are captured in
the outer disk. Its value depends critically on the velocity
distribution of neutrons and the size of the halo.
For the optimistic value $\beta \sim 1/2$ we have 
$L(\mbox{2.2 MeV})\sim 10^{-4} \dot{M} c^2$ that can be as high as 
$10^{34}$ erg/s for $\dot{M}$ close to $\dot{M}_{\rm max}$. 
This is detectable by the future INTEGRAL mission 
from a distance of up to several kpc.

\section{Conclusions}

We suggest and analyze in detail the possibility  of self-sustained
neutron halo formation in the vicinity of a disk-accreting black hole
(or a neutron star). The neutrons come from collisional dissociation of
He in the infalling matter. The term ``self-sustained'' means here
that, even if  the  ion temperature in the disk is below the
threshold  for He dissociation by protons, the neutron production is
supported by collisions of energetic neutrons from the halo with
helium nuclei.

Our analysis shows that halo formation is possible when the mass
accretion rate is below some critical value. In this case neutrons
accrete slower than ions, which leads to the neutron pile-up: the
density of neutrons can be much higher than proton density.  Neutron
halo exists in a broad range of disk parameters and can lead to the
number of observational and dynamical effects including
self-regulation of accretion and long-term variability of the
sources, enhanced hard X-ray and gamma-ray luminosity (also due to proton-neutron bremsstrahlung), 2.2 MeV
deuterium gamma-ray line resulting from neutron capture.

\end{document}